# Electrical breakdown of a dielectric for the formation of a superconducting nanocontact

S.I. Bondarenko, A.V. Krevsun, V.P. Koverya, A.G. Sivakov, R.S. Galushkov

B. Verkin Institute for Low Temperature Physics and Engineering of the National Academy of Sciences of Ukraine, 47 Nauky ave., Kharkiv, 61103, Ukraine

E-mail: bondarenko@ilt.kharkov.ua

Electrical breakdown of the dielectric nanolayer between film electrodes of niobium and an alloy of 50% indium and 50% tin forms a bridge of this alloy between the electrodes. The bridge resistance depends on the breakdown current. The length of the bridge is equal to the thickness of the dielectric (30 nm), and its diameter is 25 nm. The calculated coherence length of the alloy at 0 K is close to the length of the bridge. The calculated critical current of a bridge with a resistance of 1 Ω at a temperature of 0 K is 2 mA. It is concluded that such a bridge should have the properties of a Josephson contact at a temperature lower than the critical temperature of the alloy (6.5 K).

## 1. Introduction.

An important element of all electronic devices are electrical contacts. Numerous scientific studies and reviews are devoted to the formation and properties of contacts [1-6]. As a rule, most of them are formed by normal (non-superconducting) metals and operate at a temperature not lower than -50 $^{0}$C (223 K). In particular, one of the types of such contacts is formed during electrical breakdown (fritting) of the dielectric layer between two electrodes. In the second half of the 20th century, new, different from traditional types of contacts, superconducting contacts with a pronounced quantum properties appeared. These include the Giaever contact, in which, at helium temperatures, single electrons tunnel through the dielectric barrier between two superconductors, and the Josephson contact, in which two electrons (Cooper pairs) tunnel without dessipation and simultaneously [7] through an even thinner dielectric layer. The greatest application in quantum electronics has received Josephson tunnel contacts (JTC), on the basis of which quantum information (qubits) and magnetometric systems are developed [8,9]. Following JTC, other varieties of JTC were discovered in the form of flat film metal superconducting bridges between two wide films, some properties of which are close to those of JTC [10]. Finally, a superconducting bridge was created for the first time by electrical breakdown of a dielectric layer between two superconducting films of niobium and lead [11]. It was shown that such a bridge has Josephson properties and that a superconducting quantum interferometer can be created on its basis. At the same time, a number of its properties have not been studied. These include: the dependence of the normal resistance of the bridge on the breakdown current, the dependence of the breakdown voltage on the shape and material of the electrodes, the value of the diameter of the bridge, the effect of the dielectric material on the resistance of the bridge. The purpose of this paper is to determine the indicated properties of nanosized bridges when using niobium oxide with a thickness of 30 nm as a dielectric and film electrodes made of niobium and of a low-melting indium-tin alloy.

2. Nanobridge design.

Figure 1 shows a schematic representation of a film structure with a nanobridge in the central part of the figure.

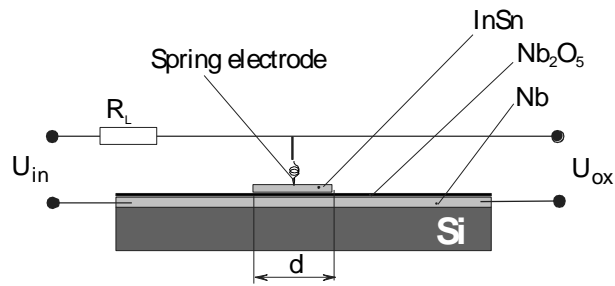

Fig.1 Schematic representation of the cross section of the experimental sample of the structure, consisting of the upper electrode in the form of a film disk with a diameter $d$ from an indium-tin alloy, a layer of niobium oxide, and a lower electrode in the form of a niobium film on a silicon substrate.

The figure also shows a constant regulated voltage ($U_{in}$) circuit on film electrodes made of niobium (anode) and 50% indium-50% tin alloy (cathode) with limiting resistance ($R_L$) and a voltage ($U_{ox}$) measurement circuit on them. The voltage for the breakdown of the dielectric is applied to the alloy film with the help of a metal spiral needle tinned with the same alloy. This design excludes mechanical damage to the dielectric nanolayer before its breakdown. Another type of film structure is shown in Fig.2. It differs in that the voltage ($U_{in}$) is applied directly to the dielectric using a niobium needle.

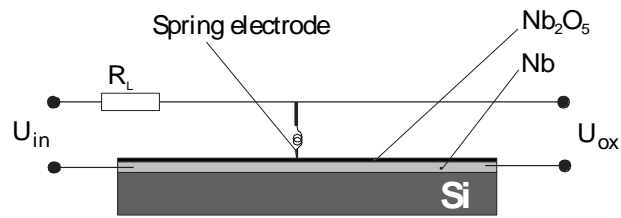

Fig.2 Schematic representation of the section of the experimental sample of the structure, consisting of the upper electrode in the form of a spring (spiral)-loaded niobium needle, a layer of niobium oxide, the lower electrode in the form of a niobium film on a silicon substrate.

### 3. Manufacturing technology of nanobridges.

The process of manufacturing nanobridges (NBs) consists of several stages:

- chemical cleaning of the niobium film located on the silicon substrate,
- formation of a dielectric in the form of niobium oxide on the surface of the niobium film,
- application of an indium-tin alloy film on the oxide in the form of a disk with a diameter of 0.5 to 0.9 mm,
- formation of two wire electrical leads from the niobium film,
- fixing the resulting film sandwich on a holder with a movable voltage-carrying electrode (Fig. 3), which also has two wires,
- connection of four leads from the film and from the current-carrying electrode to a constant voltage source through a limiting resistance and to a voltmeter to measure the voltage on the dielectric,
- mechanical movement of the voltage-carrying electrode either to the alloy film or to the dielectric up to contact with their surface and the formation of a voltage measurement circuit on the dielectric,
- a smooth increase in the voltage of the voltage generator up to the occurrence of an electrical breakdown of the dielectric with the formation of nanobridge.

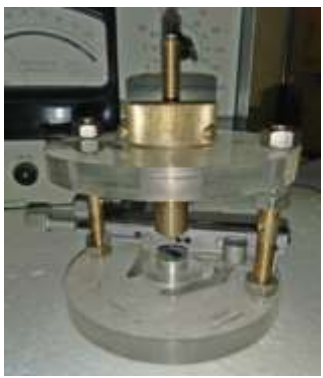

Fig.3 Outward looking attachment for micro-contacting of the movable needle with the surface of the structure during electrical breakdown of the dielectric.

Chemical cleaning of the niobium film is carried out using acidic and alkaline solutions, followed by washing with distilled water. Niobium oxide $Nb_2O_5$ with a thickness of 30 nm is formed by the electrochemical method [8] in a 1% sulfuric acid solution at a potential difference between the cathode (lead) and the anode (niobium film) equal to 10 V. An alloy film with a thickness of 100 nm is deposited by thermal evaporation of alloy portions of 50% indium and 50% tin in a vacuum chamber of a VUP-5M unit at a residual pressure of $10^{-14}$ Pa.

4. Setting up experiments.

According to the theory of breakdown of a thin dielectric layer [4], the length of the resulting cylindrical metal bridge is equal to its thickness, and its diameter is less than the length. In our case, both geometric dimensions of the bridge correspond to the generally accepted concept of nanoscale (from 1 to 100 nm). In accordance with the purpose of research, it is required at room temperature (20 $^0$C) to determine:
- the magnitude of the breakdown voltage of the selected dielectric and its dependence on the size and type of material of the film electrodes,
- dependence of the normal resistance ($R_1$) of the nanobridge in the structure "niobium film- oxide-alloy film- spiral electrode" on the electrical parameters of the breakdown.

To solve these problems, samples of the structures were made and a micrograph of one of which is shown in Fig. 4.

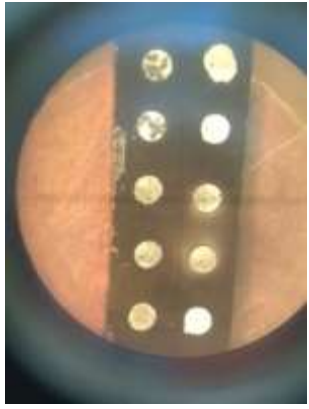

Fig. 4 External view of the experimental sample with ten indium-tin film disks with a diameter of 0.9 mm.

The figure shows 10 indium-tin film disks and the oxide surface between them. The conductive movable holder electrode for supplying voltage to the disks or directly to the oxide was made in two versions. The first of these was a spiral of copper wire coated with an indium-tin alloy. The second was a spiral of thin niobium wire. The spiral non-rigid shape of the electrode makes it possible to reduce the mechanical pressure on the oxide during contact with it and to avoid its mechanical destruction.

To determine the magnitude of the breakdown voltage and its dependence on the size of the film disk and the type of material of the cathode electrode, the breakdown was carried out at different diameters of the film disks (0.5 and 0.9 mm) and when voltage was applied to the disk or when voltage was applied to the oxide using a niobium spring (spiral) electrode. To determine the dependence of the nanobridges resistance ($R_1$) in the pressure electrode circuit on the breakdown parameters, breakdowns of dielectric between the niobium film and the alloy film were made at different values of limiting resistance ($R_L$) in the breakdown circuit (from 100 kΩ to 500 Ω). The value $R_1$ was determined by the formula

$$R_1 = U_{ox} / I_{bd} = (U_{ox} / U_{in\ bd}) R_L \qquad (1)$$

where $U_{ox}$, $I_{bd}$, $U_{inbd}$ are, respectively, a voltage on the oxide after breakdown, a current of the breakdown, the input voltage of the breakdown. The dependence of the voltage change on the electrodes in the process of increasing the voltage of the generator is shown in Fig.5. The results of the experiments were compared in the calculation formula for the resistance [4].

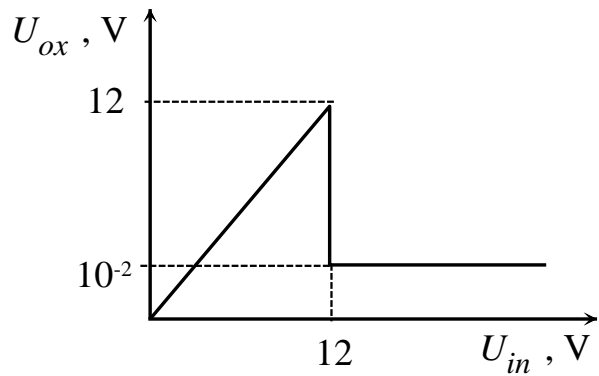

Fig.5 Dependence of voltage on the dielectric on generator voltage.

## 5. Measurement results.

As a result of measurements of the breakdown voltage on numerous samples of the experimental structure, it was found that it practically does not depend on the size of the film cathode and on the type of material from which it is made. In this case, the breakdown voltage is in the range of 12–13 V. Thus, the breakdown electric field strength of niobium oxide with a thickness of 30 nm between the alloy and niobium films was about $3 \times 10^6$ V/cm.

Measurements of the resistance of the bridges showed that it strongly depends on the magnitude of the limiting resistance in the breakdown circuit and, consequently, on the breakdown current ($I_{bd}$). The value of the resistances of the bridges turned out to be significantly less than the value of the limiting resistances. This suggests that the magnitude of the breakdown current is determined mainly by the magnitude of the limiting resistance. The nature of changes in the resistance of the bridges can be seen in Fig. 6, which shows the resistance values of 30 bridges from three groups, which differ in breakdowns at three different values of limiting resistance (100 kΩ, 10 kΩ, 1 kΩ).

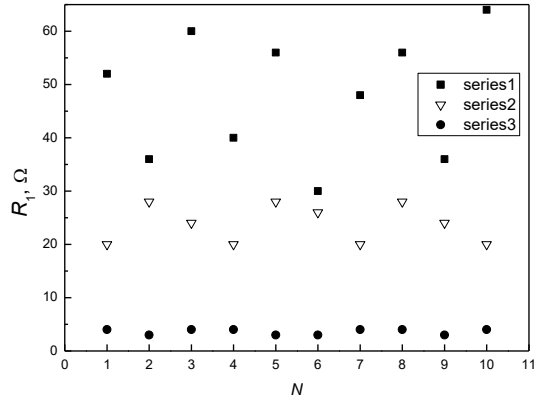

Fig.6. The spread of resistance values $R_1$ of three series of bridges with ten bridges ($N=10$) in each series obtained as a result of the breakdown of an oxide 30 nm thick between a niobium film and an indium-tin alloy film at three values of the limiting resistance $R_L$: series1 – at $R_L=100$ kΩ; series2 – at $R_L=10$ kΩ; series3 – at $R_L=1$ kΩ.

The figure shows that at high limiting resistances and, accordingly, lower breakdown currents, the resistances of the bridges increase (with a decrease in the breakdown current) and vice versa, with an increase in the breakdown current, they decrease and reach a value of 3-4 Ω. The measured resistance ($R_1$) in the pressure electrode circuit consists of the resistance of the bridge itself, the contact resistance of the spiral electrode with the alloy film, and the resistance of the electrode itself. The bridge resistance R was determined using a separate experiment, in which film leads were deposited on an alloy disk and a four-probe circuit was created to measure the bridge resistance. Measurements using this circuit showed that the bridge resistance was close to 1 Ω.

In this case, the spread of bridge resistance values from sample to sample also decreases with decreasing limiting resistance. A further decrease in the limiting resistance below 1 kΩ and, accordingly, an increase in the breakdown current above 10 mA leads to an increase in the resistance of the bridges and even to the partial destruction of film disks.

## 6. Analysis and discussion of measurement results.

The independence of the breakdown voltage from the size and material of the cathode can be explained by the fact that the relatively small diameters (0.1–2 mm) of the cathode materials used in the work (niobium and indium-tin alloy) have little effect on the distribution of field lines and the intensity of the electric field in the dielectric nanolayer between the cathode and a much larger anode in the form of a rectangular (12 mm × 4 mm) niobium film. As a result, the electric field strength sufficient to breakdown the dielectric is determined only by the thickness and material of the dielectric (assuming that the density of defects in the dielectric remains approximately the same in all samples). This assumption is confirmed by our preliminary measurements of the breakdown voltage of similar samples with a 1.5 times greater thickness of niobium oxide (45 nm). The breakdown voltage increased by a factor of 1.5, which corresponds to the maintenance of the breakdown electric field strength of niobium oxide at a level of about $3 \times 10^6$ V/cm.

The resistance formed during the breakdown of metal bridges between metal electrodes arises under the action of a strong electric field (more than $10^6$ V/cm) in the dielectric as a result of emission of electrons from the cathode to the anode [9]. The emission causes the formation of an electron avalanche in a narrow channel between the electrodes with a current density of more than $10^6$ A/cm$^2$ [10], which causes the destruction of the dielectric and melting of the electrodes. First of all, this happens with the electrode having the lowest melting point. In our case, this is an alloy film electrode. During this fast process, the electrode metal enters the breakdown channel [4], cools to solidification and forms a metal bridge of the cathode material with possible inclusions of the dielectric material. This is confirmed by our earlier studies of samples in which the cathode was made of lead and the critical temperature of the bridge was equal to the critical temperature of lead [11].

Based on the obtained dependence of the bridge resistance on the breakdown current, it can be assumed that cleaner bridges with lower resistance are formed due to a higher breakdown current density. In order to check the purity of the formed bridge with a resistance of about 1 Ω, an experiment was carried out on the comparative measurement of the specific resistance of the bridge $\rho_b$ and the alloy film $\rho$. According to the theory [4], the ratio of these resistances is equal to the inverse ratio ($k$) of their thermal resistance coefficients ($\alpha_b$ and $\alpha$):

$$k = \rho_b / \rho \approx \alpha / \alpha_b \geq 1 \quad . \tag{2}$$

The values $k$, $\rho_b$ were determined by measuring $\rho$, $\alpha$, $\alpha_b$. It was found that $k \approx 1$.

Thus, it was proved that the purity of the material of such a nanobridge is close to the purity of the alloy film. In this case, the resistance of the bridge can be approximately described by formula [4]:

$$R \approx \rho(ks / \pi a2 + 1/ a) , \tag{3}$$

where $a$ is the bridge diameter. Using this formula, knowing $\rho$, $k$, $s$, one can estimate the value of the bridge diameter $a$. It is convenient to make this assessment using the graphs of the dependences of the bridge resistance on its diameter for various values of $k$ (Fig. 7):

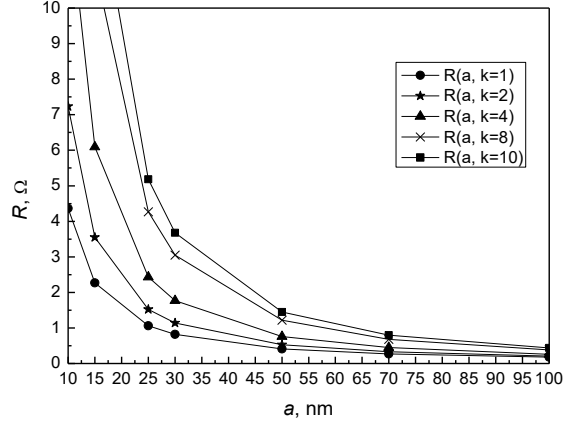

Fig.7 A series of dependences of the resistance ( R ) of the bridges on the diameter of the bridge (*a*), the thickness of the dielectric (*s* = 30 nm), the ratio of the temperature coefficients of the resistivity (*k*).

We can see from the graphs that at $s$=30 nm, $k$ =1, $\rho \approx 10^{-5}$ Ω cm and $R$=1 Ω, the bridge diameter is 25 nm.

To evaluate the Josephson properties of a superconducting nanobridge with such parameters, it is necessary to compare its length with the coherence length of an alloy with 50% indium and 50% tin. The critical magnetic field of this alloy at a temperature of 0 K is $B_c$=0.27 T [15]. We determined the critical temperature of the alloy is about 6.5 K, and the penetration depth of the magnetic field ($\lambda$) at a temperature of 4.2 K is about 60 nm [16]. For a superconductor of the second kind, which is an alloy, according to the indicated data, it is possible to approximately determine the coherence length ($\xi$) of the bridge material at $T$= 0 K using the formula [17]:

$$B_c = \Phi_0 / (\lambda \xi), \qquad (4)$$

where $\Phi_0$ is the quantum of the magnetic flux. After substituting the values, we obtain that $\xi(0) \approx 25$ nm. When NB is used at liquid helium temperature (4.2 K), the coherence length of the bridge will increase and become close to its length. Thus, it can be assumed that the Josephson properties of the NB will be ensured.

Using Silsby's rule [18], one can estimate the critical current $I_c$(0 K) of such a bridge:

$$I_c(0K) = a H_c / 4. \qquad (5)$$

If we take into account that $H_c$ (0 K)=2.7×10$^5$ A/m, then $I_c$(0 K)≈2×10$^{-3}$ A.

# 7. Summary.

As a result of measuring the electrical breakdown voltage of a dielectric in the form of niobium oxide with a thickness of 30 nm, separating two electrodes in the form of a niobium film (anode) and an indium-tin alloy film (cathode), and the analysis of the measurement results of these samples, it was found that with a smooth increase voltage on the electrodes, breakdown occurs at a voltage of 12-13 V, which is practically independent of the size and material of the cathode. The strength of the electric field that causes the breakdown of niobium oxide at room temperature is about $3 \times 10^6$ V/cm and is determined by the thickness and dielectric properties of the oxide.

A strong influence of the breakdown current, determined by the limiting resistance in the breakdown circuit, on the electrical resistance of the metal bridge that appears between the electrodes after breakdown has been discovered and studied. It has been established that bridges with the lowest resistance (1-3 Ω) are formed at a breakdown current of 10 mA. This value of resistance is characterized by the greatest repeatability, which is confirmed by the breakdown of dozens of samples.

The analysis of the calculation formula for determining the diameter of the bridge, based on the known values of the resistance of the bridge, its length and the ratio between the coefficients of thermal resistance (CTR) of the bridge and the cathode material (indium-tin alloy film) is carried out. This ratio determines the purity of the bridge material relative to the cathode material. A family of dependences of the resistance on the bridge diameter is obtained for various values of the ratio between the CTR. As a result of the experimental determination of the CTR, using the obtained dependences, the diameter ($a$) of a bridge with a resistance of about 1 Ω ($a = 25$ nm) and the ratio between the CTS of the bridge and the cathode film were determined. This ratio turned out to be close to unity. This ratio indicates that the bridge with a length of 30 nm, obtained as a result of the breakdown of the niobium oxide nanolayer, practically consists of the same alloy from which the cathode is made.

The coherence length of a nanobridge made of an alloy of 50% indium and 50% tin was calculated at $T=0$ K. It turned out to be close to the nanobridge length. This allows us to assume that in the superconducting state ($T<T_c=6.5$ K), a nanobridge with a resistance of 1 Ω will have Josephson properties.

Based on the diameter of the bridge and the value of the critical field of the indium-tin alloy, an estimate was made of the critical current ($I_c$) of the nanobridge at $T= 0$ K ($I_c \approx 2 \times 10^{-3}$ A).